\documentclass[prd,aps,12pt,nofootinbib,reprint,showkeys]{revtex4-1} 
\usepackage{mathrsfs,amssymb,amsmath,amsfonts,amsxtra}
\usepackage{graphicx,color,verbatim}
\usepackage[colorlinks]{hyperref}
\usepackage[utf8x]{inputenc}

\begin{document}
\sloppypar \sloppy
\title{The Wallstrom objection as a possibility to augment quantum theory}
\author{I. Schmelzer}
\thanks{Berlin, Germany}
\noaffiliation

\begin{abstract}
Wallstrom has argued that quantum interpretations which construct the wave function starting from Madelung variables \(\psi(q)=\rho(q)\exp(\frac{i}{\hbar}S(q))\), in particular many variants of Nelsonian stochastics, are not equivalent to quantum theory. Accepting this, we explicitly add the physical restriction \((\forall q\in Q)|\psi(q)|^2>0\) in the configuration space representation to quantum theory. The resulting theories depend on the choice of the configuration space \(Q\). 

The restriction adds possibilities to falsify such theories,  making them preferable following Popper's criterion of empirical content until they have been really falsified. 

Considering the case of a relativistic scalar field, we argue that it is reasonable to expect that the variant based on particle ontology can be (or even has been already) falsified, while to falsify the variant based on field ontology seems much harder, if not impossible.

If correct, switching to a field ontology seems a reasonable choice for interpretations based on Madelung variables to circumvent the Wallstrom objection. 
\end{abstract}

\maketitle


\newcommand{\pd}{\partial} 
\newcommand{\ud}{\mathrm{d}} 
\newcommand{\f}{\varphi}
\newcommand{\w}{\nabla\times v}
\renewcommand{\a}{\alpha}

\renewcommand{\H}{\mbox{$\mathcal{H}$}} 
\newcommand{\B}{\mbox{$\mathbb{Z}_2$}}
\newcommand{\Z}{\mbox{$\mathbb{Z}$}}
\newcommand{\R}{\mbox{$\mathbb{R}$}}
\newcommand{\C}{\mbox{$\mathbb{C}$}}

\newtheorem{theorem}{Theorem}
\newtheorem{principle}{Principle}
\newtheorem{postulate}{Postulate}
\newcommand{\Sch}{Schr\"{o}dinger\/ }

\section{Introduction}

There is a whole class of interpretations of quantum theory which is based on Madelung variables\footnote{These variables have been originally proposed by Madelung \cite{Madelung} for one-particle \Sch theory, with the aim to give it a hydrodynamic interpretation. As a consequence, they are usually named ``hydrodynamic variables''. For the actual applications in the interpretation of quantum theory a hydrodynamic interpretation makes no sense, given that the theory is defined on the configuration space, not the physical space. To name them ``hydrodynamic'' is therefore grossly misleading, so that I name them here (and propose to do this in general) ``Madelung variables''.} -- the fields \(\rho(q), S(q)\) defining the wave function by \(\psi(q) = \sqrt{\rho(q)}\exp(\frac{i}{\hbar}S(q))\) \cite{Madelung}-\cite{Caticha}. They can be used if the Hamiltonian is quadratic in the momentum variables. As the typical example for such a theory, one usually starts with non-relativistic single-particle theory:
\begin{equation}
H = \frac{1}{2m} \langle p,p\rangle + V(q).
\end{equation}
Then, the \Sch equation in these variables gives a continuity equation for \(\rho(q)\)
\begin{equation}\label{continuity}
\partial_t \rho(q) + \nabla(\rho(q) \vec{v}(q)) = 0
\end{equation}
with a velocity defined by the gradient of \(S(q)\):
\begin{equation}
m\vec{v}(q) = \nabla S(q).
\end{equation}
together with a quantum generalization of the classical Hamilton-Jacobi equation for $S(q)$
\begin{equation}\label{Bohm}
-\pd_t S(q) = \frac{1}{2m} \langle\nabla S(q), \nabla S(q)\rangle + V(q) + Q(q).
\end{equation}
which differs from the classical Hamilton-Jacobi equation only by a so-called quantum potential:
\begin{equation}
Q(q) = -\frac{\hbar^2}{2m} \frac{\Delta \sqrt{\rho}}{\sqrt{\rho}}.
\end{equation}
To generalize this to many-particle theory is straightforward. Less well-known is that relativistic field theory can also made fit into this scheme -- which destroys the popular prejudice that such interpretations can cover only non-relativistic theories.

Madelung variables are, without doubt, helpful if one wants to consider the classical limit \(\hbar\to 0\). In this limit, the \Sch equation automatically gives the classical Hamilton-Jacobi equation together with a corresponding continuity equation for the probability flow. But what we are interested here is another application of these variables, namely in interpretations where the variables \(\rho(q),S(q)\) together with the equations \eqref{continuity}, \eqref{Bohm} are considered as fundamental, while the wave function \(\psi(q)\) together with the \Sch equation are constructed out of them. This is the base of many variants of Nelsonian stochastics \cite{Nelson}-\cite{Caticha}.

This is unproblematic only in one direction: every valid global solution of \eqref{continuity}, \eqref{Bohm} defines a valid solution of the \Sch equation. Unfortunately, not every solution of the \Sch equation can be constructed in this way, but only solutions with \(|\psi(q)|^2 > 0\) everywhere. As argued by Wallstrom \cite{Wallstrom1,Wallstrom2}, near the zeros of the wave function the equivalence fails, so that theories starting from Madelung variables give theories which are not equivalent to quantum theory -- they forbid zeros of the wave function.

A simple but uninteresting solution is to give up the idea to derive the \Sch equation from \eqref{continuity}, \eqref{Bohm}. This seems popular among proponents of de Broglie-Bohm theory: While Madelung variables have been used by Bohm himself \cite{Bohm}, they don't play a central role in many modern presentations of dBB theory like \cite{Duerr}, \cite{Duerr2}. Given that the equations of dBB theory anyway have to postulated instead of being derived, this is not a big loss. The situation is different for Nelsonian stochastics and its variants, where \eqref{continuity}, \eqref{Bohm} can be derived, giving them a much superior status in comparison with the \Sch equation itself.

One way to solve this problem would be to handle the zeros of the wave function differently. This requires already an explicit modification of quantum theory itself, because near the zeros of the wave function \(|\vec{v}|\) becomes infinite and, moreover, the flow is no longer a potential (irrotational) flow globally. The resulting theory would have to be a different, subquantum theory. In \cite{againstWallstrom}, the author has considered such alternatives and found that they could plausibly in a quantum limit lead to the condition \(0 < \Delta \rho(q) < \infty\) almost everywhere at the zeros of the wave function. This would allow zeros of the wave function in general position, so that arbitrary close to every quantum solution where would be a solution which could be obtained in this way.

This solution requires the acceptance that quantum theory itself has to be modified, and survives near the zeros of the wave function only as an approximation. An even more serious problem would be that in the more fundamental theory there exists no wave function and no potential of the flow \(S(q)\). For some interpretations based on Madelung variables this may be unproblematic. But in particular for Caticha's entropic dynamics \cite{CatichaBook,Caticha} this would be fatal, because the function \(S(q) + \ln \sqrt{\rho(q)}\) is interpreted there as an entropy. This allows a derivation of the equations from first principles of entropic inference. This derivation from first principles could not be preserved in a subquantum theory following \cite{againstWallstrom}, because there would not exist a potential \(S(q)\) for the velocity in the more fundamental subquantum theory.

In this paper, we evaluate another possibility, namely a quite radical, straightforward one -- to add the condition that \(|\psi(q)|^2 > 0\) holds for all $q\in Q$ to quantum theory. This defines a new theory, with Nelsonian stochastics as well as its many variants as possible interpretations. It would be empirically falsifiable by the preparation of states which would have to have zeros in the configuration space representation. Given that this would not falsify quantum theory, this gives additional empirical content, thus, makes it superior to quantum theory (following Popper's criterion of empirical content) until it is actually falsified.

The condition \(|\psi(q)|^2 > 0\) holds only for the wave function in the configuration space $Q$. Zeros in, say, the momentum representation would be irrelevant and could not falsify the new theory. So, theories with different choices of the configuration space -- a choice which seemed to be purely metaphysical up to now -- define now physically different theories. 

We will argue below that the choice of $Q$ does really matter, by considering two popular choices of the configuration space for the case of a scalar field theory: We will name ``stochastic particle theory'' the theory based on the particle ontology, where the number and position of particles defines the configuraion, and ``stochastic field theory'' the theory based on the field ontology, where the field values $\varphi(x)$ in space define the configuration. 

Generalizations to field theories with non-zero spin we leave to future research, given that, on the one hand, there will be known problems defining both variants for the same spin (positions for photons, as well as classical field configurations for fermions). On the other hand, for the conceptual problems considered in this paper, a rough look suggests that these problems will not play a decisive role. 

\section{General problems of preparing states with zeros}

Let's consider now what we have to do to construct a state which violates the condition \(|\psi(q)|^2 > 0\) for some $q\in Q$.

One naive hope could be that the evolution would do everything itself: Say, if we would start with an arbitrary wave function, possibly without zeros, the evolution would automatically create, in some time, some zeros. This does not work. Instead, it appears that the quantum potential term in equation \eqref{Bohm} gives a good protection against the density becoming zero. Indeed, for \(|\psi(q)|^2 \to 0\) the quantum potential obtains large negative values, \(Q(q)\to -\infty\) at least at the minimum of \(|\psi(q)|^2\). This creates an inflow into this minimum. So, at least with the physical level of certainty, we can be sure that a probability distribution which is nonzero everywhere initially remains nonzero everywhere forever.

Another naive idea would be to construct a sufficiently large number of basic states, so that all other states could be obtained as a superposition of those basic states. Unfortunately in the stochastic theories we do not have a superposition principle: If there exists states in Madelung variables leading to the wave functions \(\psi_1(q), \psi_2(q)\), there may be no Madelung variables for their superpositions \(\alpha_1\psi_1 + \alpha_2\psi_2\). That means, superpositions simply may not exist. So, we have to create the problematic state itself, and have to do this in a pure form.

This creates a problem with creating states of moving particles. How do we create a moving particle state? A moving particle will be emitted at some origin and then hit some target. After this we can even make a good guess when it was emitted. But what we can create is essentially only the emitter state. This state of the emitter, with some probability, emits the particle. But the result is not a pure state of a flying particle. During the time of possible emission, we have only a superposition of a state of an emitted particle and a not yet emitted particle. The ``not yet emitted particle'' part is the vacuum state of the corresponding field, which is a ground state with no zeros at all. So, the construction by constructing the emitter would give (at every particular place) only a superposition of the state with a flying particle with a sufficiently large vacuum contribution. Once the vacuum does not have any zeroes, a large enough vacuum contribution gives not much hope to create states with unavoidable zeroes in this way.

So, we would better restrict our considerations to stationary states. 

But even in this case, some other contributions cannot be avoided. To see this we have to consider here the way how to prepare particular states in general. This general preparation procedure of a quantum state starts with the quantum system being in an unknown state, measures some operator \(\hat{\alpha}\) with eigenstates \(\psi_i(q_{sys})\), so that the measurement result $i$ prepares the state in that eigenstate  \(\psi_i(q_{sys})\). So we start with 
\[\psi_{sys}(q_{sys})= \sum_i \alpha_i \psi^{sys}_i(q_{sys})\]
with unknown values $\alpha_i$. We combine it with a measurement device, which is assumed to be in its ground state \(\psi_0(q_{dev})\), which we can be assumed to be without zeroes. We also have to assume both are independent, so that the wave function of the whole system is \(\psi = \psi_{sys}(q_{sys})\psi_0(q_{dev})\). Now we let the whole system interact in a way which measures the operator \(\hat{\alpha}\). The \Sch evolution gives after the interaction
\[ \psi(q_{sys},q_{dev}) = \sum_i \alpha_i \psi^{sys}_i(q_{sys}) \psi^{dev}_i(q_{dev}).\]
Now the question appears how the measurement result is identified. Here, we have to follow the prescriptions of the realist interpretations under consideration. They all have definite trajectories of the measurement device  \(q_{dev}(t)\) which describe the actual state of the device, and they all assume also that it is this actual state \(q_{dev}(t)\) which is observed as the measurement result of a macroscopic, classical device. 

As a consequence, all these interpretations share also to rule how to define the resulting state of the quantum system: We have to put the observed value of the classical measurement result  \(q_{dev}(t)\) into the full wave function \(\psi(q_{sys},q_{dev})\) to obtain a wave function \(\psi_{sys}(q_{sys})\) of the quantum system:
\[ \psi_{sys}(q_{sys}) = \sum_i \alpha_i \psi_i(q_{sys}) \psi^{dev}_i(q^{obs}_{dev}).\]
In an ideal experiment, the eigenfunctions \(\psi^{dev}_i(.)\) would not overlap at all, and the value  $q^{obs}_{dev}$ would identify the measurement result exactly, and therefore the result would be one of the eigenstates \(\psi^{sys}_i(q_{sys})\), namely the only one where \(\psi^{dev}_i(q^{obs}_{dev})\neq 0\), with \(\psi^{dev}_j(q^{obs}_{dev})= 0\) for  all \(j\neq i\). But, again, to obtain what stochastic mechanics predicts we have to follow the rules of stochastic mechanics, and in stochastic mechanics we have no exact zeros of the wave function. So, all what we can hope for is that \( |\psi_j(q^{obs}_{dev})|^2 \ll 1\) for all \(j\neq i\).
As a consequence, we will not be able to construct in any real experiment pure eigenstates of interesting operators. All we can hope for is 
\[ \psi_{sys}(q_{sys}) \approx \psi^{sys}_i(q_{sys}). \]
So, all that we can construct is only some  minor modification \(\psi(q) + \delta\psi(q)\) with some small but unknown error contribution \(\delta\psi(q)\). 

Thus, a zero of the wave function has to be stable against minor modifications of the wave function, else we could not claim that we have constructed a state with a zero of the wave function. 

Last but not least, let's note that to prepare a particular state in a way which is sufficiently certain to identify a zero, one has to restrict oneself to very low energy. This is quite obvious, given that states with higher energies will usually somehow move, but we have to restrict ourselves to stable states. Unfortunately, the state with the lowest energy -- the vacuum -- is one without any zeros. Fortunately, for many stablee states with low energy it remains possible to create and identify them correctly. But if the energy increases, this becomes much harder, and this difficulty increases very fast. 

\section{Particle ontology: Angular momentum eigenstates of particles}

Fortunately, such minor distortions are not sufficient to get rid of all zeroes completely. While for a one-dimensional configuration a small distortion of the wave function is always sufficient to get rid of zeros, in two dimensions we can already have zeros in a general position, that means, adding an arbitrary small distortion will only shift the position of the zero, but does not allow to get rid of it completely. 

The classical example of such states are angular momentum eigenstates with nontrivial angular momentum in one-particle non-relativistic quantum theory. One would not expect that relativistic effects become relevant here, so that we will ignore relativistic effects.\footnote{Relativistic approaches with particle ontology exist, in particular following \cite{Bohm53}, and they are quite popular in the context of dBB theory \cite{Duerr2}. But to consider here and now relativisitic corrections does not seem necessary, given that what has to be constructed are anyway only stable states, and that the argument depends only on quite rough qualitative properties of these states.} It is this example which is mentioned in the original papers by Wallstrom \cite{Wallstrom1},\cite{Wallstrom2}. 

So, let's consider a state with $L_z=1$ and restrict ourselves to the plane $z=0$. Then the zero is located at $x=y=0$ and surrounded by a circle $\sqrt{x^2+y^2} = r_0$ where $|\psi|^2 =\rho_0>0$ and the dependence on the angle $\varphi$ is $e^{i\varphi}$. Let's consider a small deformation \(\psi(\vec{x})\to\psi(\vec{x}) + \delta\psi(\vec{x})\). If $|\delta\psi|^2 \ll \rho_0$,  the deformation will be insufficient to lead to a zero at $r_0$. Thus, during the whole deformation the image of the circle $r_0$ will remain inside $\mathbb{C}^*$, its winding number cannot change and has to remain $1$. And therefore there has to remain a zero of the wave function inside the circle too.

Are more objections possible? Yes. The states with small but nonzero angular momentum eigenstates one would start to think about would be (ignoring spin) states of electrons in atoms. But in everyday life these atoms themselves move around. Thus, the same general objections against moving states can be applied here too. To meet this objection, one would have to fix, additionally, also the position of the atomic nucleus. Moreover, it has to be fixed in a sufficiently accurate way so that the additional uncertainty created by the position of the nucleus does not distort the state of the electron so much that the proof that there has to be a zero inside fails. This would, if realized, falsify the version of the theory that the wave function does not have zeros in the configuration space if the configuration space is defined by a particle ontology. 

\subsection{Falsification based on quantum computation?}

Another already existing candidate would be the constructions of quantum computers. One can reasonably guess that there exist various states of a quantum computer which have stable zeros in the sense described above. This is a question which would have to be clarified by specialists for quantum computation. 

In principle, one would have to study the consequences of using stochastic particle theory for quantum computation too. Last but not least, quantum computation depends on quantum theory being accurate and completely unrestricted by conditions that there should be no zeros of the wave function in a particular representation. Would quantum computation fail in stochastic particle theory?  We cannot exclude this possibility. Of course, the equations are exactly the same in stochastic particle theory as in quantum theory. So, from this point of view, nothing changes. But the quantum computer starts with a preparation procedure, and, as shown above, such a preparation procedure will not give pure non-vacuum eigenstates, but only states with small distortions. If these small distortions prevent the preparations of states with zeros initially, the future evolution will prevent them during the whole computation. But what if intermediate states with such zeros play a key role in the computation itself? Can this possibly create systematic errors of quantum computers in some unavoidable way? 

This question has to be left for future reasearch. But, independent of this, the possibility that the construction of quantum computers has already created states which can be, with sufficient certainty, identified as pure states with zeros which are stable against minor distortions, is quite plausible. 

Thus, it seems quite plausible that stochastic particle theory can be easily falsified. Probably existing constructions of quantum computers already contain states which would falsify the particle ontology -- all what would be necessary is to evaluate such constructions from this point of view.  If not, then one would expect that to construct such states is possible, and all what has to be done is to realize such a construction.

\section{Field ontology: The case of relativistic scalar field theory}

Let's now consider an example with a field ontology. While the field ontology defines a much more natural candidate in particular for relativistic field theories, it has been largely ignored in considerations about hidden variable theories. Indeed, it is quite common to claim that such hidden variable theories like dBB as well as Nelsonian stochastics exist only for non-relativistic particle theory. But in fact the general scheme works nicely for all Hamiltonians with a quadratic dependence on the momentum variables, and in relativistic field theories we have this type of dependence. So, for relativistic scalar field theory, we have

\begin{equation}
 H = \frac12 \int \pi(x)^2 + \nabla \varphi \nabla \varphi + m^2\varphi^2 d^3x.
\end{equation}
In this paper, we restrict ourselves to the case of a single scalar field.\footnote{For various approaches to handle gauge fields in stochastic field theory see \cite{Namsrai}. Note that such  approaches are not obliged to implement modern BRST quantization schemes, they can even simply break gauge invariance so that they can be handled like vector fields. Fermions can be obtained using a \(\mathbb{Z}_2\)-valued field theory, which can be obtained as the low energy sector of a scalar field theory with a degenerated vacuum state. Following \cite{clm}, a \(\mathbb{Z}_2\)-valued lattice theory in a three-dimensional space gives in the large distance limit a doublet of Dirac fermions, which would be sufficient to describe all fermions of the SM with massive neutrinos.} 

In this case the configuration space is defined by the space of functions on usual space \(Q\cong\{ \varphi(.):\mathbb{R}^3\to\mathbb{R}\}\). To avoid the various infinities related with field theories, one can, instead, use a lattice regularization on a large enough cube with periodic boundary conditions, which gives the finite dimensional configuration space \(Q\cong\{ \varphi(.):\mathbb{Z}_N^3\to\mathbb{R}\}\). This regularization distorts essentially nothing relevant below. 

Using the same considerations as used above for the particle ontology, we find similarly that we have to restrict ourselves to stationary low energy states.

But in the field ontology, a stationary single particle state is described in a way which differs a lot from the case of particle ontology. A stationary state is classically a standing wave, thus, has the form $\varphi(x,t) = f(t)\varphi(x)$ and therefore depends only on a single coordinate of the configuration space, a linear combination of the natural coordinates $q^x$ defined by the value $
\varphi(x)$ of the field at the point $x$, so that $q^{\varphi} = \int \varphi(x) q^x$ is the coordinate associated with a particular standing wave field configuration $\varphi(x)$. The wave functional depends only on this particular coordinate, $\Psi(q) = \Psi(q^{\varphi}$. 

Unfortunately, in a one-dimensional space one cannot construct any state where the wave function has zeros which cannot be removed by a small distortion. As in one-dimensional quantum theory, all the non-vacuum eigenstates will be real and have zeros, and once we have a symmetry $\varphi \to -\varphi$, for the one-particle eigenstate the zero will be one at the origin, $\Psi(0)=0$. But this zero will not be stable, the minor imaginary contribution of the vacuum $\Psi(q^\varphi)\to \Psi(q^\varphi) + i\varepsilon \Psi_{vac}(q)$ will destroy it.

A stable zero requires at least a two-dimensional subspace. In principle, a moving particle would give it. But this leads to the problem with states of moving particles considered above: What can be resonably constructed is the state of an emitter, and correspondingly only a superposition of the vacuum with the particle states emitted at different times. Thus, it seems quite plausible that in this case the vacuum contribution will be large enough at every particular point to prevent zeros. 

\section{Conclusions}

We have studied the possibility to meet the Wallstrom objection against Nelsonian stochastics and similar stochastic interpretations of quantum mechanics in a quite radical way: To embrace it by explicitly modifying quantum theory, adding a new postulate to quantum theory that the wave function in the configuration space representation does not have any zeros. 

Because the modified theory can be empirically falsified without falsifying quantum theory itself, the modification adds empirical (predictive) power to quantum theory, and therefore has to be preferred by Popper's criterion of empirical content in comparison to standard quantum theory until it is empirically falsified.  

Given that the additional predictions depend on the choice of the configuration space, variants of stochastic mechanics based on different choices of the configuration space become different physical theories.  

We have considered two examples of such theories for the case of a scalar field theory, namely the variants with particle ontology (stochastic particle theory) and with field ontology (stochastic field theory).

We have identified some problems related with possible empirical falsifications of such theories. The main problem is that these theories have no superposition principle -- one cannot be sure that a superposition of valid pure states of the theory is also a valid pure state. So, if one wants to interpret some experiment as a falsification of such a theory, one has to be careful that one does not use, even implicitly, a superposition principle somewhere in the interpretation of the experiment. Moreover, preparation procedures have uncontrollable uncertainties, so that one has to construct wave functions with zeros which are stable against minor modifications. 

For the particle ontology, it seems nonetheless plausible that the construction of states with non-trivial angular momentum allows to falsify the theory. With some probability, what has been already reached for the construction of quantum computers may be already sufficient to falsify stochastic particle theory. Even if not, nothing seems to endanger the possibility that states which would falsify the particle ontology could be constructed. 

The consideration of field ontology suggests a different result. Stationary one-particle states appear insufficient to prove the existence of a stable zero in a wave function. It seems not implausible that more complex states, which have zeroes even if distorted by minor modifications of the wave function, will not be stationary states. This would make it difficult to construct them in a sufficiently certain way. 

Thus, it is at least not implausible that the final result would be an empirical falsification of stochastic mechanics for particle ontology together with the failure to falsify in a similar way stochastic mechanics for field ontology.

If this guess survives the consideration of fields with spin and real experiments will be left to future research. 

If it appears correct, the Wallstrom objection appears to be not a bug but a tool to show that stochastic field theory is even superior to quantum field theory, given its superiority according to Popper's criterion of empirical content and the failure to falsify it empirically. 

On the other hand, it leads to an empirical argument against a particle ontology of field theories. Thus, the classical ``particle or wave'' question of quantum theory is no longer metaphysical, but allows an empirical answer. 

Of course, this empirical answer is of limited value, because what can be falsified is only stochastic particle theory, not quantum particle theory, thus, proponents of particle theory have a natural fallback theory. 

On the other hand, there are anyway already strong metaphysical arguments against a particle ontology for field theories. In particular, the very definition of the particles depend, in semiclassical gravity, on the particular state of the gravitational field. While this is unproblematic for pseudoparticles like phonons, but it would be strange for fundamental particles. Moreover, for a photon, there exists not even a position measurement operator (see, for example, \cite{Bacry}). Thus, a field ontology seems preferable anyway. In this situation, to add an empirical argument in favor of a field ontology could be decisive.

Last but not least, let's note that this paper shows that what looks metaphysical may soon, and in completely unsuspected ways, become physical. The way how it has happened here is not untypical: There was a problem of a particular metaphysical interpretation. To solve this problem, it appears natural and necessary to modify the theory. This shows that the consideration of different interpretations should not be rejected as metaphysical -- it is a natural step toward the development of different physical theories, theories which have to be created to solve particular metaphysical problems of a particular interpretation.

\end{document}